\documentclass[]{aa} 
\usepackage{subfigure}                                   
\usepackage[dvipsnames]{color}

\usepackage{graphicx}
\usepackage{txfonts}
\begin{document}
 \title{Do we expect to detect electromagnetic radiation from merging stellar mass black binaries like GW150914? No.}

   \author{Shuang-Nan Zhang\inst{1,2}, Yuan Liu\inst{1}, Shuxu Yi\inst{1}, Zigao Dai\inst{3,4}, and Chaoguang Huang\inst{6}
          }

   \institute{$^1$ Key Laboratory of Particle Astrophysics, Institute of High
Energy Physics, Chinese Academy of Sciences, Beijing
100049, China\\
              \email{zhangsn@ihep.ac.cn}
              \\$^2$ National Astronomical Observatories, Chinese Academy of Sciences, Beijing 100012, PR China
              \\$^3$ School of Astronomy and Space Science, Nanjing University, Nanjing 210093, China
              \\$^4$ Modern Astronomy and Astrophysics, Nanjing University, Ministry of Education, Nanjing 210093, China
               \\$^5$ Division for Theoretical Physics, Institute of High
Energy Physics, Chinese Academy of Sciences,Beijing 100049, China
         }

   \date{...}


   \abstract
   {The LIGO consortium announced the first direct detection of gravitation wave event GW150914 from two merging black holes; however the nature of the black holes are still not clear.}
   {We study whether electromagnetic radiation can be detected from merging stellar mass black binaries like GW150914.}
   {We briefly investigate the possible growth and merging processes of the two stellar mass black holes in the merging event of GW150914 detected by aLIGO, as clocked by a distant external observer.}
   {Our main results are: (1) The description of the black hole growth using stationary metric of a pre-existing black hole predicts strong electromagnetic radiation from merging black holes, which is inconsistent with GW150914;  (2) Only gravitational wave radiation can be produced in the coalescence of two black holes such as that in the GW150914 event, if the black hole growth is described using time-dependent metric considering the influence of the in-falling matter onto a pre-existing black hole, as clocked by a distant external observer.}
   {Future high sensitivity detections of gravitational waves from merging black holes might be used to probe matter distribution and space-time geometry in the vicinity of the horizon. Perhaps the GW150914-like events can be identified with traditional astronomy observations only if the black holes are embedded in extremely dense medium before their final merge, when very strong electromagnetic radiation is produced and can escape from the system.}


 \authorrunning{S.-N. Zhang et al.}
   \titlerunning{How to identify BH mergers like GW150914}

   \maketitle
%

\section{Introduction}
As a prediction of general relativity, gravitational wave (GW) will open a new window for astronomy observations. Besides the direct detection of a GW event, identifying its counterpart in electromagnetic (EM) band is also of importance in understanding the full physical process underlying the GW event. The Advanced Laser Interferometer Gravitational-Wave Observatory (aLIGO) detected a
 GW signal GW150914 on September 14, 2015 09:50:45 UTC (The LIGO Scientific Collaboration \& the Virgo Collaboration 2016). This GW event is explained as originated  from two stellar mass black holes (SmBHs) with masses of $36^{+5}_{-4} M_\odot$ and $29^{+4}_{-4} M_\odot$ (in the source frame). No evident counterpart is identified in follow-up observations of X-ray, optical, or UV band, except for a possible weak  transient source reported by the Fermi/GBM team (Connaughton et al. 2016).

 A SmBH is commonly believed to be formed as a result of stellar evolution, which produces a SmBH with masses several times that of the Sun, as evidenced by the mass distributions of the SmBHs in X-ray binaries (Zhang 2013). However, the masses of the two SmBHs found in GW150914 are much larger than the masses of SmBHs in all known X-ray binaries. Therefore, they should have significantly grown by accretion after their births, e.g., from the companion star, the fall-back gas, or the interstellar medium. As emphasized by Vachaspati et al. (2007), the observational phenomena of the growth of a BH should and can only be discussed in the frame of a stationary observer outside the BH, because the ``proper time" experienced by the in-falling comoving observer (CO) can not be mapped back to the time of the external observer (EO) once the CO crosses the event horizon (EH) of the BH. Therefore whenever and whatever the CO experiences and ``sees" is irrelevant to the EO, once the CO reaches and then crosses the EH.

 In this work, we first study the possible EM emission in the conventionally used stationary metric of BHs when two SmBHs merge. We then study what happens when considering the time-dependent metric when matter falls onto BHs. Finally we discuss how future high sensitivity detections of gravitational waves from merging black holes might be used to probe matter distribution and space-time geometry in the vicinity of the horizon..

 \section{EM radiation of merging SmBHs in stationary metric}

As emphasized by Vachaspati et al. (2007), it will take infinite time  (coordinate time) for the in-falling matter to arrive at the EH of a Schwarzschild BH as clocked by an EO; this is a fundamental property of the EH, as first pointed out in the seminar work by Oppenheimer and Snyner (1939). Actually the last statement in the abstract of Oppenheimer and Snyner (1939) is ``an external observer sees the star asymptotically shrinking to its gravitational radius".  In other words, within the limited time of the EO, the in-falling matter never really arrives at the EH and certainly cannot cross the EH, and therefore must accumulate just outside the EH of the BH. This is the well-known ``frozen star'' or ``black star'' phenomenon  and widely presented in textbooks  (Misner et al. 1973, Schutz 1985, Shapiro \& Teukolsky 1983, Weinberg 1972).  The above conclusion is based on the (widely-adopted) simplification that metric of the BH is stationary, i..e, it is a background metric and not influenced by the in-falling matter; in other words, the in-falling matter is taken as a test particle.

Vachaspati et al. (2007) considered the quantum effect and argued that the ``pre-Hawking radiation'' can evaporate all in-falling matter in the collapsing process. Thus, the BH will not grow and it may resolve the black hole information loss problem. However, the time scale of such evaporation process is much longer than the Hubble time for any macro astrophysical object (Vachaspati 2007; Greenwood et al. 2011). Hence, realistically it is completely negligible in the growth process of any astrophysical BHs; the discovery of the SmBH merging event GW150914 is a clear evidence against any significant evaporation process during the significant growth of the two SmBHs.

Alternatively, Vachaspati (2007) pointed out that significant EM radiation is expected in the merging process of two such ``black stars'' with the accumulated matter outside their EHs. Thus, the coalescence  of such ``black stars''  could be empirically distinguished from the coalescence  of two \textit{bona fide} BHs (i.e. all mass is inside the EH). However, Petrovay (2007) argued that it will also take infinite time to observe such radiation by a distant EO and the radiation will be gravitationally redshifted and diluted  into unobservable level, and thus observationally it is still impossible to distinguish ``black stars'' from \textit{bona fide} BHs, if only considering the EM radiation from their merging process.

However, during the in-falling process, the angular momentum of the in-falling matter may compress and amplify magnetic fields around the EH of a SmBH or  ``black star'', which may produce jet by a BZ-like mechanism (Blandford \& Znajek 1977) and avoid the infinite redshift to produce significant EM radiation, if the SmBH is rapidly spinning. Therefore, even before the merging of the two SmBHs, significant EM radiation may be observable. During and after the merging process, more intensive EM radiation should be released, since the SmBH must be spinning and all the previously accumulated matter must be accreted onto it.

In fact, almost all of the accumulated matter could form as a transient torus due to its high angular momentum. In the following, we estimate the mass of this accretion disk. The BZ luminosity of a jet producing a weak short gamma-ray burst (GRB) observed by Fermi/GBM is approximated by (Lee et al. 2000)
\begin{equation}
L_{\rm BZ}=6.0\times 10^{51}B_{{\rm BH},14}^2M_{{\rm BH},60}^2a^2f(a)\,{\rm erg}\,{\rm s}^{-1},
\end{equation}
where $B_{\rm BH}=B_{{\rm BH},14}\times 10^{14}$~G is the inner magnetic field strength of the accretion disk, $M_{\rm BH}=M_{{\rm BH},60}\times 60M_\odot$ is the mass of the post-merger BH, $a\simeq 0.67$ is the rotation parameter of the post-merger BH, and $f(a)=1-[(1+(1-a^2)^{1/2})/2]^{1/2}\simeq 0.067$. The inner magnetic energy density of the accretion disk can be estimated by the pressure near the horizon,
\begin{equation}
B_{\rm BH}^2/8\pi\simeq P_{\rm in}.
\end{equation}
Xue et al. (2013) estimated this pressure through
\begin{equation}
{\rm log}(P_{\rm in}/{\rm erg}\,{\rm cm}^{-3})\approx 27.4+1.22a+1.00{\rm log}\dot{m}-2{\rm log}M_{{\rm BH},60},
\end{equation}
where $\dot{M}=\dot{m}M_\odot$\,s$^{-1}$ is the accretion rate and the viscosity parameter $\alpha$ is taken as a constant of $0.1$ (also see Liu et al. 2015). Thus, the BZ luminosity is calculated by
\begin{equation}
L_{\rm BH}\simeq 7.5\times 10^{51}\dot{m}\,{\rm erg}\,{\rm s}^{-1}.
\end{equation}
Please note that the BZ luminosity derived here is independent of the BH mass. This luminosity could be collimated with an opening angle of $\theta_{\rm j}$, and should be related with the observed isotropic radiation luminosity of the short GRB through
\begin{equation}
L_{\rm BH}=\epsilon_\gamma^{-1}L_{\gamma,{\rm iso}}(1-\cos\theta_{\rm j}),
\end{equation}
where $\epsilon_\gamma$ is the prompt radiation efficiency of the short GRB and $L_{\gamma,{\rm iso}}$ is  the observed isotropic radiation luminosity. Because the INTEGRAL team reported non-detection of any significant gamma-ray radiation from GW150914 (Savchenko et al. 2016), we can take the reported Fermi/GBM detection as an upper limit, i.e.  $L_{\gamma,{\rm iso}}\leq 1.8\times 10^{49}\,{\rm erg}\,{\rm s}^{-1}$ (Connaughton et al. 2016). Therefore, we can derive the accretion rate
\begin{equation}
\dot{m}\simeq 10^{-4}(\epsilon_\gamma/0.1)^{-1}(\theta_{\rm j}/0.1)^2,
\end{equation}
and the total mass of the accretion disk
\begin{equation}
M_{\rm torus}\leq 10^{-4}M_\odot(\epsilon_\gamma/0.1)^{-1}(\theta_{\rm j}/0.1)^2(\tau_\gamma/1\,{\rm s}),
\end{equation}
where $\tau_\gamma$ is the duration of the short GRB. This mass is significantly larger than the one estimated by Li et al. (2016) and Yamazaki et al. (2016). We suggest that such a torus with $M_{\rm torus}$ could have been produced due to fall-back of some ejecting materials during the supernova explosion of one pre-merging BH and could have become unstable due to tidal disruption during the merger of two BHs. The other possible ways by which this torus forms have been discussed by Loeb (2016) and Perna et al. (2016).

The above estimated mass $M_{\rm torus}\leq 10^{-4}M_\odot$ accumulated around the BH is significantly smaller than that expected if the two SmBHs were grown up from an initial mass of around $10M_\odot$. On the other hand, if around $20M_\odot$ of matter is indeed accumulated during their growth process, the expected gamma-ray luminosity through the BZ process should be around $L_{\gamma,{\rm iso}}\sim 10^{55}\,{\rm erg}\,{\rm s}^{-1}$, which can be easily detected by essentially all previous GRB instruments even if the events originated from the very early universe.

Since GW150914 is a nearby event and it is expected such events should happen very frequently in the whole universe, non-detection of such bright GRBs over the past several decades completely rejects the expectation obtained using the stationary metric. This conclusion is generic and independent of where the infalling matter comes from and how the SmBHs were formed, as far as they were formed by astrophysical collapse. This is because in the original Oppenheimer and Snyner (1939) calculation of direct collapse of a massive star, essentially all mass of the final BH is outside its EH, as clocked by a distant external observer. In this case we will have $L_{\gamma,{\rm iso}}\gg 10^{55}\,{\rm erg}\,{\rm s}^{-1}$ for the GW150914-like events.

 \section{The case for time-dependent metric}

However, solving self-consistently the Einstein's field equation for the whole gravitating system of an in-falling thick shell of matter and a Schwarzschild BH, Liu \& Zhang (2009) found that the metric must be time-dependent and the shell can cross the expanding EH even clocked by a distant EO within finite time, except for the outer boundary of the shell. Furthermore, if the existence of the environmental material is also considered, all of the in-falling matter can cross the final EH within finite time clocked by the EO. Thus, \textit{bona fide} BHs can form even in the frame of a distant EO; BHs formed this way are called astrophysical BHs here, since all known BHs found so far with astronomical observations are formed by collapse of matter.

To demonstrate this effect, we follow Liu \& Zhang (2009) to calculate an idealized and simple case: two shells of matter free-fall onto a 10 $M_\odot$  BH; the inner shell has 1 $M_\odot$  and outer shell has 0.1 $M_\odot$, starting the in-falling process from distances of 20$r_0$ and 25$r_0$, respectively, where $r_0=2GM/c^2$ and $M$ is the total mass of the two shells and the BH. The initial thickness of each shell is 2$r_0$. The inner shell and outer shells mimic the actual in-falling matter onto the BH (i.e. the growth process of a SmBH in the GW150914 event) and the residual accretion from the environmental material surrounding the system, respectively. The complete in-falling process as clocked by a distant EO is shown in Figure 1. The inner shell crosses the EH in a very short time, i.e., $8\times10^{-3}$ s clocked by the EO.

\begin{figure}
	\centering
	\includegraphics[width=7cm]{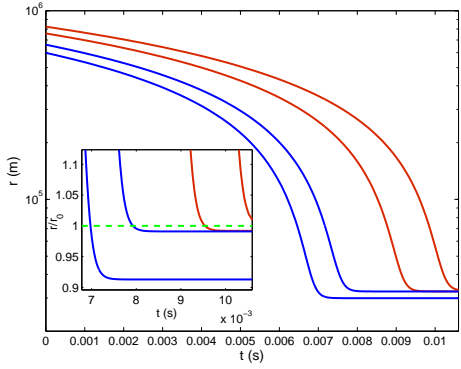}
	\caption{The in-falling process of two shells (inner shell mass 1 $M_\odot$  and outer shell mass 0.1 $M_\odot$) towards a pre-existing 10 $M_\odot$ SmBH, as clocked by a distant external observer. The inset shows that the complete inner shell crosses the event horizon of the final SmBH with 11.1$M_\odot$ within $8\times10^{-3}$ s. }
\end{figure}

More realistically, the in-falling process can be considered as two stages. In the first stage, the matter falls via an accretion disk; in the second stage, the matter is nearly free-fall after the innermost stable circular orbit, qualitatively similar to that shown in Figure 1. Therefore the total time scale of the growth of BH is determined completely by the first stage. Despite of different accretion modes, we estimate the total time scale of the growth of BH by the typical accretion rate of an X-ray binary, i.e., $\sim10^{-8} M_\odot$. Therefore, it will take $\sim10^9$ yr for a typical BH in an X-ray binary ($\sim10 M_\odot$) to growth to $\sim30 M_\odot$ . However, it is possible that the two SmBHs have been grown up in much denser environment with much higher accretion rate, $\sim10^9$ yr can be considered a rough upper limit for their ages.

The time to merger for a circular black hole binary with radius $a$ and component masses $m_1, m_2$ is given by Peters (1964), if the decay of the orbit driven by energy and angular momentum loss resulting from the emission of gravitational waves:
\begin{equation}
\tau_\textrm{GW} = \frac{5}{256} \frac{c^5}{G^3} \frac{a^4}{m_1 m_2 (m_1+m_2)}\, .
\end{equation}
The coalescence time needed for two SmBHs in the GW150914 event as a function of the binary orbital radius is shown in Figure 2, if the only orbital energy loss mechanism is through releasing GW energy. Mandel \& de Mink (2016) showed that BH binaries like that in the GW150914 event can be formed through chemically homogeneous evolution in short-period stellar binaries and $\tau_\textrm{GW}$ is typically in the range of 4 to 11 Gyr after formation, corresponding to an orbital radius of around 0.1 AU.  Alternatively,
Belczynski et al. (1016) showed that within their classical evolutionary scenario (Belczynski et al. 2010a, b), the stellar progenitors of the BHs constituting GW150914 most likely formed in low metallicity environments in the early universe and thus the SmBHs in GW150914 should have experienced coalescence time at least several Gyr before the final merger. Therefore, the coalescence time is much longer than the growth time scale of the EH, both clocked by the same EO. This means, long before the final merging stage of the two SmBHs in the GW150914 event, practically all matter should have disappeared into the two SmBHs. As a natural result, only gravitational wave radiation can be produced in the coalescence of two astrophysical BHs such as that the GW150914 event; this was explicitly predicted in Liu \& Zhang (2009).

This conclusion is also generic and independent of where the in-falling matter comes from and how the BHs were formed, because in any physically feasible scenario the in-falling matter will disappear into the expanding EH within time scales much shorter than any other time scales involved in producing the final merging event, as far as they were formed through astrophysical collapse.

\begin{figure}
	\centering
\includegraphics[width=7cm]{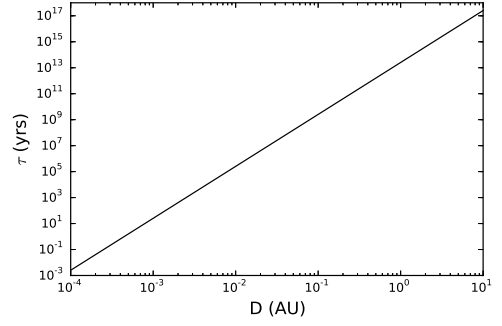}
	\caption{The coalescence time needed for two SmBHs in the GW150914 event as a function of the binary orbital radius.}
\end{figure}

\section{Summary, conclusion and discussion}

We briefly investigated the possible growth and merging processes of the two stellar mass black holes in the merging event of GW150914 detected by aLIGO, as clocked by a distant external observer but using two different metric systems.

In the conventional way of describing the black hole growth process through matter falling into a seed black hole using the stationary metric of this black hole, all in-falling matter must be accumulated outside the event horizon of the black hole, within the finite time clocked by the distant external observer. When two such black holes merge, the accumulated matter will inevitably produce strong electromagnetic radiation. However, the direct electromagnetic radiation can not escape from the system to the distant external observer, due to the infinite redshift caused by the event horizon of the black hole.

On the other hand, the compressed matter can significantly amplify the large scale magnetic field near the event horizon of the black hole and allow the spinning black hole to produce strong electromagnetic radiation escaping to the distant external observer. Taking the Fermi/GBM reported GRB luminosity as an upper limit to the purported GRB associated with GW150914, we found that the mass accumulated outside the black hole event horizon is $\sim 10^{-5}M_\odot$ and significantly smaller than that expected if the two black holes in GW150915 were grown up from an initial mass of around $10M_\odot$. However, if around $20M_\odot$ of matter is indeed accumulated during their growth process to make up the $\sim 30M_\odot$ for each of the black holes, the expected gamma-ray luminosity should be $\sim 10^{55}\,{\rm erg}\,{\rm s}^{-1}$, which have never been observed but can be easily detected by essentially all previous GRB instruments even if the events originated from the very early universe..

Using the time-dependent metric by considering the in-falling matter and the pre-existing black hole, we find that practically all in-falling matter can free-fall into the event horizon of the final black hole within very short time, also clocked by a distant external observer. Even considering the formation of an accretion disk around the black hole, the total accretion time scale is still much shorter than the coalescence time needed for the two black holes in the GW150914 event for any reasonable binary orbital radius, if the only orbital energy loss mechanism is through releasing GW energy. Therefore, long before the final merging stage of the two black holes in the GW150914 event, practically all matter should have disappeared into the two black holes.

Our main conclusions are:
\begin{enumerate}
\item The description of the black hole growth using stationary metric of a pre-existing black hole is inconsistent with GW150914.

\item Only gravitational wave radiation can be produced in the coalescence of two black holes such as that in the GW150914 event.

\end{enumerate}

Both conclusions are generic and independent of where the in-falling matter comes from and how the BHs were formed, as far as they were consequences of astrophysical collapse and/or accretion.

In stationary metric, it is usually anticipated that matter around the BHs should fall down toward the horizon and approach infinitely, as discussed in section 2. However,  it may be possible that for some reason (e.g., due to modification to gravity, or radiation pressure, or significant angular momentum of the in-falling matter) the matter can stay outside the horizon at a finite distance. In the time-dependent metric, the matter is anticipated also not at $r=0$, but instead distributed within the BHs. In both cases, such matter can participate in the tidal interaction between the merging BHs during the inspiral process and modify the gravitational waveform.

In the formal case (i.e., stationary metric), tidal interaction can arise from the (presumably rather weak) reflection of GWs by matter that stays outside the EH, as discussed by Li and Lovelace (2008).  More dramatically, if the BH turns out to be a ``gravastar'' (as proposed by Mazur and Mottolla 2004) which does not posses a horizon at all, GWs that were to go down the horizon may even resonate within the gravastar, as discussed by Pani et al.~(2010). The impact of such matter to the ringdown waveform has recently been discussed by generally Barausse, Cardoso and Pani (2014), and in the context of GW150914 by Cardoso, Franzin and Pani (2016) and Chirenti and Rezzolla (2016).

Therefore future high sensitivity detections of gravitational waves from merging black holes might be used to distinguish between astrophysical BHs and mathematical singularities (i.e., all their gravitational masses are concentrated at the location of $r=0$), and even to probe matter distribution and space-time geometry in the vicinity of the horizon, the impacts of which to the outgoing gravitational waveforms need to be studied more systematically, possibly with numerical general relativity simulations.

Finally, we suggest that the GW150914-like events can be identified with traditional astronomy observations only if the black holes are embedded in extremely dense medium before their final merge, when very strong electromagnetic radiation is produced and can escape from the system (e.g., Murase, K. et al. 2016). However, in this case, the dense medium should accelerate the coalescence process of the stellar mass black hole binary, such that the waveform of the gravitational waves may deviate from the black hole merging in vacuum; numerical general relativity simulations should be made to distinguish these two possibilities.


\begin{acknowledgements}
We thank Prof. Yanbei Chen, who can not be a co-author due to some restrictions as a member of the LIGO collaboration, for extensive discussions and actual contribution to this manuscript. This work is partially supported by the National Basic Research Program (``973'' Program)
of China (Grants 2014CB845802 and 2012CB821801), the National Natural Science
Foundation of China (grants No. 11103019, 11133002, 11103022, and 11373036), the Qianren start-up grant 292012312D1117210, and the Strategic Priority Research Program
``The Emergence of Cosmological Structures'' (Grant No. XDB09000000) of the Chinese Academy
of Sciences.
\end{acknowledgements}

\end{document}